\documentclass[a4paper,11pt]{article}
\usepackage{amsthm}
\usepackage{amsmath}
\usepackage{amssymb}
\usepackage{color}
\usepackage{cite}
\usepackage{graphicx}
\usepackage{xcolor, slashed}
\usepackage{subfigure}
\usepackage[linktocpage]{hyperref}
\usepackage{setspace}
\usepackage{comment}

\usepackage{titletoc}
\dottedcontents{section}[1.2em]{}{1em}{0.5pc}
\dottedcontents{subsection}[4em]{}{2em}{1pc}

\usepackage{etoolbox}
\makeatletter
\pretocmd{\subsection}{\addtocontents{toc}{\protect\addvspace{-4\p@}}}{}{}
\pretocmd{\section}{\addtocontents{toc}{\protect\addvspace{-4\p@}}}{}{}
\makeatother
\newcommand{\mrd}{\mathrm{d}}
\renewcommand{\H}{{_\mathrm{H}}}
\newcommand{\M}{{_{M}}}
\renewcommand{\S}{{_{S}}}
\newcommand{\Ji}{{_{{J}_i}}}

\definecolor{darkred}{rgb}{0.9,0.05,0.05}
\definecolor{darkblue}{rgb}{0.05,0.05,0.6}
\definecolor{darkgreen}{rgb}{0.05,0.6,0.05}
\definecolor{brightgreen}{rgb}{0.1,0.9,0.1}

\renewcommand*{\eqref}[1]{%
	\begingroup
	\hypersetup{
		linkcolor=red,
		linkbordercolor=darkred,
	}%
	\textcolor{red}{(\ref{#1})}%
	\endgroup 
}

\hypersetup{linkcolor=red,citecolor=brightgreen,urlcolor=blue,colorlinks=true}

\begin{document}
	\let\endtitlepage\relax
	
	\begin{titlepage}
		\begin{center}
			\renewcommand{\baselinestretch}{1.5}  
			\setstretch{1.5}
			
			\vspace*{-0.5cm}
			
			{
				{\Large{\bf{Gauge Invariant Derivation of Zeroth and First Laws of\\ Black Hole Thermodynamics}}}}
			
			\vspace{9mm}
			\renewcommand{\baselinestretch}{1}  
			\setstretch{1}
			
			\centerline{\large{Kamal Hajian$^{\dagger\ast}$}\footnote{khajian@metu.edu.tr, kamalhajian@uni-oldenburg.de}, \large{M. M. Sheikh-Jabbari$^\ddagger$}\footnote{jabbari@theory.ipm.ac.ir }, \large{Bayram Tekin$^\dagger$}\footnote{btekin@metu.edu.tr}}

			\vspace{5mm}
			\normalsize
			\textit{$^\dagger$Department of Physics, Middle East Technical University, 06800, Ankara, Turkey}\\
			\textit{$^\ast$Institute of Physics, University of Oldenburg, Postfach 2503, D-26111 Oldenburg, Germany}\\
			\textit{$^\ddagger$School of Physics, Institute for Research in Fundamental Sciences (IPM),  P.O.Box 19395-5531, Tehran, Iran}
			\vspace{5mm}
			
			\begin{abstract}
				In gauge invariant theories, like Einstein-Maxwell theory, physical observables should be  gauge invariant. In particular, mass, entropy,  angular momentum, electric charge and their respective chemical potentials, temperature, horizon angular velocity and electric potential which appear in the laws of black hole thermodynamics should  be gauge invariant.  In the usual construction of the laws of black hole thermodynamics, gauge invariance of the intensive quantities is subtle; here we remedy this and provide a gauge invariant derivation and the proof of the zeroth and first laws of black hole thermodynamics.
			\end{abstract}
		\end{center}
		
	\end{titlepage}

	\renewcommand{\baselinestretch}{1.05}  
	\setstretch{1}
	\section{Introduction}
	
	Symmetries have been the main guiding principle in formulating laws of physics. They  help us  identify conserved quantities, through the celebrated Noether's theorem, as well as formulate interactions, through the notion of local gauge symmetries.  Maxwell's theory was the first such example which was realized to be a gauge theory with the $U(1)$ gauge symmetry. This symmetry appeared to be the key to formulating quantum electrodynamics (QED). An extension of the same notion to other gauge groups laid the foundation of what then led to the standard model of particle physics. In a parallel line of development, Einstein's equivalence principle was formulated through diffeomorphism invariance of general relativity (GR), which is a local gauge symmetry under general coordinate transformations. Gauge symmetries introduce nonpropagating and generically nonphysical degrees of freedom into the description; nonetheless all physical observables ought to be gauge invariant.   
	
	Defining (conserved) charges  has been one of the outstanding questions in the context of GR, as the theory is not only diffeomorphism invariant, but also background independent. There have been several proposals to do so starting with Komar \cite{Komar:1958wp} in 1958, followed by Arnowitt-Deser-Misner \cite{Arnowitt:1959ah,Arnowitt:1960es,Arnowitt:1962hi},  Regge–Teitelboim Hamiltonian method \cite{Regge:1974zd}, Brown-York \cite{Brown:1992br}, the ADT formulation \cite{Abbott:1981ff,Deser:2002rt,Deser:2002jk}, and more recently \cite{Barnich:2001jy, Barnich:2007bf, Chen:2021szm}, to just name a few. Each of these formulations have their own advantages and disadvantages, see chapter 5 of \cite{The-BH-Book} for a recent review. In particular, the covariant phase space formalism (CPSF), which was developed by many in particular by Wald and collaborators \cite{Lee:1990gr,Iyer:1994ys,Wald:1999wa,Ashtekar:1987hia,Ashtekar:1990gc,Crnkovic:1987at}, has provided a suitable framework to compute diffeomorphism invariant conserved charges associated with spacetime symmetries. 
	
	Black holes are ubiquitous solutions to theories of gravity and have raised many questions at conceptual, theoretical, and observational level since the conception of Einstein's GR in 1915. After seminal works of Bekenstein, Hawking, and many others since early 1970s, it is now well accepted that black holes satisfy laws of thermodynamics. The idealized stationary black holes are classically in thermal equilibrium at a given temperature with an entropy and other charges and associated chemical potentials. There have been many works on proving laws of black hole thermodynamics from the first principles, starting from diffeomorphism invariant theories of gravity. The first steps were taken by Hawking in proving the  constancy of surface gravity on the horizon of a stationary black hole as the zeroth law and the area theorem as a statement of the second law; see \cite{The-BH-Book} for a detailed discussion and review. The next breakthroughs came by the seminal works of Wald, where it was shown that the entropy is the conserved charge \'a la Noether, associated with a Killing vector generating the horizon \cite{Wald:1993nt}, and then a derivation of the first law was given as a result of diffeomorphism invariance of the theory \cite{Iyer:1994ys}. 
	
	Wald's derivations were based on CPSF, and despite the elegance and depth, the original analysis had some loose ends; e.g., it relied  on notions of mass and angular momentum at asymptotically flat spacetimes, and the electric charge was not included. Some of these points were improved and addressed in later works. In particular, it was noted in \cite{HS:2015xlp} that for stationary black holes with a Killing horizon, one can relax the dependence on the asymptotic behavior of the spacetime geometry and define the first law at the horizon or any other codimension 2 slice surrounding the horizon. 
	
	In this work, we address another loose end of Wald's derivation: It is not manifestly  gauge invariant. The early steps in this direction were taken in \cite{Gao:2001ut, Gao:2003ys, HS:2015xlp, Prabhu:2015vua, Compere:2007, Elgood:2020svt}; see also \cite{The-BH-Book}. In this work we provide a full proof of gauge invariance of each and every thermodynamical quantity appearing in the first and zeroth laws of black hole thermodynamics.

	\section{The setup and conventions}
	
	Consider a generic $D$ dimensional gravitational scalar-Maxwell theory described by the action
	\begin{equation}\label{action}
		I=\int \mrd ^D x \sqrt{-g} \left(\mathcal{L}_{_0}(g_{\mu\nu},\phi_a, \dots) + \frac{1}{4}\sigma(\phi_a)F_{\mu\nu}F^{\mu\nu}  \right)=\int \big(\mathbf{L}_{_0}+ \frac{1}{2}(\sigma \star F) \wedge F\big), \
	\end{equation}
	in which the dynamical fields are the metric $g_{\mu\nu}$, the gauge field $A$, scalar fields $\phi_a$, and ellipsis  refer to other possible dynamical fields which do not directly couple to $A$. $\mathbf{L}_{_0}=\star \mathcal{L}_{_0}$ is a generic covariant gravitational Lagrangian, and $\sigma(\phi_a)$ is an arbitrary differentiable function of some scalar fields $\phi_a$, which couples them to the electromagnetic field. In the second equality, we have adopted form notation of \cite{Iyer:1994ys}, where Lagrangian $\mathbf{L}_{_0}$ is a $D$-form, $F=\mrd A$ is the $U(1)$ gauge field strength, and $F_{\mu\nu}$ denote its components.  $\star F$ is its Hodge dual, which is a $D-2$ form. We are using the conventions in which the speed of light, Planck's constant, and Newton's constant are all set to one. These may be recovered upon dimensional analysis, if needed. For our analysis below, we need the explicit form of the gauge field equations of motion:
	\begin{equation}\label{EOM}
		\mrd (\sigma \star  F)=0.
	\end{equation}
	
	The above action is invariant under diffeomorphisms  $x^\mu\to x^\mu-\xi^\mu(x)$ as well as the  gauge transformation $A\to A+\mrd\lambda$. Under a diffeomorphism, a generic field $X$ transforms as $X\to X+ \mathcal{L}_\xi X$, and under a gauge transformation, only the gauge field transforms. We collectively denote the diffeomorphism and gauge generators by $\eta=\{\xi,\lambda\}$. 
	
	\paragraph{Charge Computation.}
	
	Covariant phase space formulation of charges  \cite{Ashtekar:1987hia,Ashtekar:1990gc,Crnkovic:1987at,Lee:1990gr,Wald:1999wa,Iyer:1994ys} is a method that facilitates our discussion on the gauge invariance of charges (see reviews  in \cite{The-BH-Book,Hajian:2015eha,Seraj:2016cym}). Here we recapitulate the basics of this method. The formulation begins with a Lagrangian density $\mathcal{L}$ in $D$ dimensions and the action $I=\int \mathbf{L}\equiv \int \star \mathcal{L}$ with some dynamical fields $\Phi(x^\mu)$. Variation of the Lagrangian reads
	\begin{equation}\label{delta L}
		\delta \mathbf{L}= \mathbf{E}_\Phi \delta \Phi + \mrd \mathbf{\Theta},    
	\end{equation}
	where $\mathbf{E}_\Phi= 0$ are the equations of motion. We will use $\approx$ to denote on-shell equality, e.g., $\delta \mathbf{L}\approx \mrd \mathbf{\Theta}$. By another variation of the symplectic potential $\mathbf{\Theta}(\delta\Phi,\Phi)$, we obtain the symplectic current $\boldsymbol{\omega}$ 
	\begin{equation}
		\boldsymbol{\omega}(\delta_1\Phi,\delta_2\Phi,\Phi)=\delta_1\mathbf{\Theta}(\delta_2\Phi,\Phi)-\delta_2\mathbf{\Theta}(\delta_1\Phi,\Phi).   
	\end{equation}
	Since $\delta^2 \mathbf{L}=0$ identically, recalling the identity $\delta \mrd = \mrd \delta $ and assuming the linearized field equations $\delta \mathbf{E}_\Phi \approx 0$, the symplectic current is locally conserved on-shell $\mrd \boldsymbol{\omega}\approx 0$ and by virtue of Poincar\'e Lemma, $\boldsymbol{\omega}\approx \mrd \boldsymbol{k}(\delta_1\Phi, \delta_2\Phi, \Phi)$.
	
	For a generic diffeomorphism and gauge transformation $\eta=\{\xi,\lambda\}$, a charge variation can be defined by
	\begin{align}
		\slashed\delta Q_\eta  &= \int_\Sigma \boldsymbol{\omega}(\delta\Phi,\delta_\eta\Phi,\Phi)= \int_\Sigma \left(\delta\mathbf{\Theta}(\delta_\eta\Phi,\Phi)-\delta_\eta\mathbf{\Theta}(\delta\Phi,\Phi)  \right) \label{del H omega}\\
		&=\oint_{\partial \Sigma}\boldsymbol{k}_\eta(\delta\Phi,\Phi)\,,  \label{del H}  
	\end{align}
	where $\Sigma$ is a $D-1$ dimensional spacelike hypersurface (a Cauchy surface), ${\partial \Sigma}$ is the $D-2$ dimensional boundary of $\Sigma$, $\delta_\eta \Phi=\{\mathcal{L}_\xi\Phi,\delta_\lambda \Phi\}$, and $\mrd\boldsymbol{k}_\eta(\delta\Phi,\Phi)\approx \boldsymbol{\omega}(\delta\Phi, \delta_\eta\Phi, \Phi)$. 
	Note that each term in a Lagrangian $\mathbf{L}=\mathbf{L}_1+\mathbf{L}_2+ \mathbf{L}_3 + \dots $ contributes to $\mathbf{\Theta}$, $\boldsymbol{\omega}$, and $\boldsymbol{k}_\eta$, in an additive way, i.e., $\mathbf{\Theta}=\mathbf{\Theta}_1+\mathbf{\Theta}_2+\mathbf{\Theta}_3 +\cdots$, $\boldsymbol{\omega}=\boldsymbol{\omega}_1+\boldsymbol{\omega}_2+\boldsymbol{\omega}_3+\cdots$, and $\boldsymbol{k}_\eta=\boldsymbol{k}_{1\eta}+\boldsymbol{k}_{2\eta}+\boldsymbol{k}_{3\eta}+ \dots$. 
	
	\paragraph{Integrability of the charge.} The charge variation $\slashed\delta Q_\eta$ is not necessarily a variation of a charge $Q_\eta$ on the phase space. The charge variation is integrable if $(\delta_1\slashed\delta_2 - \delta_2\slashed\delta_1) Q_\eta=0$. In general if the symmetry generators $\eta=\{\xi, \lambda\}$ are also field dependent, as is in the cases of our interest, this condition takes the form \cite{The-BH-Book, Geoffrey, HS:2015xlp}
	\begin{equation}\label{integrability}
		\oint_{\partial\Sigma}\ \xi\cdot \boldsymbol{\omega}(\delta_1\Phi,\delta_2\Phi,\Phi)+ \boldsymbol{k}_{\delta_1\eta}(\delta_2\Phi,\Phi)-\boldsymbol{k}_{\delta_2\eta}(\delta_1\Phi,\Phi) \approx 0.
	\end{equation}
	
	\section{Gauge noninvariance of charge}\label{app:1}
	
	While $\slashed\delta Q_\eta$ in \eqref{del H} is by construction diffeomorphism invariant ${\cal L}_{\tilde\xi}(\slashed\delta Q_\eta)=0$ for a generic $\tilde\xi$, it is not invariant under a generic gauge transformation $A\to A+\mrd \tilde{\lambda}$, $\delta_{\tilde\lambda} (\slashed\delta Q_\eta) \neq 0$. To see this, consider the Lagrangian in \eqref{action}; i.e., $\mathbf{L}=\mathbf{L}_{_0}+\mathbf{L}_A$ where $\mathbf{L}_{_0}=\mathbf{L}_{_0}(g_{\mu\nu},\phi_a,\dots)$ and $\mathbf{L}_{_A}=  \frac{1}{2}(\sigma(\phi_a)\star F)\wedge F$. For this Lagrangian, $\boldsymbol{k}_\eta=\boldsymbol{k}_{{_0}\eta}+\boldsymbol{k}_{{_\text{A}}\eta}$.  It is readily seen that $\boldsymbol{k}_{{_0}\eta}$ is invariant under the  $\tilde\lambda$ gauge transformation and we hence focus on $\boldsymbol{k}_{{_\text{A}}\eta}$. 
	Starting from $\mathbf{L}_{_A}$, its symplectic potential is found to be $\mathbf{\Theta}_{\!_A}(\delta A)=(\sigma\star F) \wedge \delta A$. Then
	\begin{equation}\label{dH--eq} 
		\begin{split}
			\boldsymbol{\omega}_{_A}(\delta, \delta_\eta) &= 
			\delta\big((\sigma \star F) \wedge \delta_\eta A \big)- \delta_\eta\big((\sigma \star F) \wedge \delta A \big) 
			\\
			&= \delta\big((\sigma \star F) \wedge (\mathcal{L}_\xi A + \mrd \lambda)\big)- \mathcal{L}_\xi\big((\sigma \star F) \wedge \delta A \big). 
		\end{split}
	\end{equation}
	In the last equation we replaced $\delta_\eta A= \mathcal{L}_\xi A + \mrd \lambda$ and used $\delta_\lambda \delta A=0$ as $\delta_\lambda( A_2-A_1)=(A_2+\mrd {\lambda})-(A_1+\mrd {\lambda})-(A_2-A_1)=0$.  Using the Cartan identity $\mathcal{L}_\xi X= \xi\cdot \mrd X + \mrd (\xi \cdot X)$ 
	for an arbitrary form $X$, we obtain
	\begin{align}
		\boldsymbol{\omega}_{_A}(\delta, \delta_\eta)&=
		\delta\big((\sigma \star F) \wedge (\xi\cdot \mrd A  +\mrd (\xi\cdot A + \lambda))\big) - \xi\cdot \mrd\big((\sigma \star F) \wedge \delta A\big)-\mrd \big(\xi \cdot \big(\sigma \star F) \wedge \delta A\big) \nonumber\\ 
		&=\mrd \left[\delta \big(\sigma \star F \wedge (\xi\cdot A+ \lambda)\big)
		-\xi \cdot \big(\sigma \star F \wedge \delta A\big)\right], \label{dH fourth eq}
	\end{align}
	where we used \eqref{EOM}, $\delta \mrd = \mrd \delta$ and $\delta\big(\sigma \star F \wedge (\xi\cdot \mrd A)\big)\approx \xi\cdot \mrd\big(\sigma \star F \wedge \delta A\big)$   to obtain \eqref{dH fourth eq}.  
	So, we find $\boldsymbol{\omega}_{_A}(\delta, \delta_\eta)=\mrd \boldsymbol{k}_{{_\text{A}}\eta}$, where
	\begin{equation}\label{kA eta}
		\boldsymbol{k}_{{_\text{A}}\eta}= \delta \big(\sigma \star F \wedge (\xi\cdot A+ \lambda)\big)-\xi \cdot \big(\sigma \star F \wedge \delta A\big).
	\end{equation}
	While the second term in \eqref{kA eta} is gauge invariant, the first term is not: $\delta_{\tilde\lambda}(\boldsymbol{k}_{{_\text{A}}\eta})= \delta \big(\sigma  (\xi\cdot \mrd\tilde\lambda) \star F \big)\neq 0$. 
	Note that the above analysis on gauge noninvariance of the charge variation also works for field-dependent gauge transformations when  $\delta \lambda\neq 0$. 
	
	Although \eqref{kA eta} is not gauge invariant for an arbitrary $\eta$, but if we augment a diffeomorphism $\xi$ by the specific gauge transformation,
	\begin{equation}\label{lambda xi.A}
		\lambda=-\xi\cdot A,
	\end{equation}
	then the first term in \eqref{kA eta} vanishes, and the result is manifestly gauge invariant. Explicitly, this happens because with this choice for $\lambda$,
	\begin{equation}\label{delta-eta-A}
		\delta_\eta A=\mathcal{L}_\xi A - \mrd (\xi\cdot A)= \xi\cdot F, \qquad F=\mrd A,
	\end{equation}
	is manifestly gauge invariant. So,  $\slashed\delta Q_\eta$ for the ``augmented diffeomorphism'' $\eta=\{\xi,-\xi\cdot A \}$ is both gauge and diffeomorphism invariant. This augmented diffeomorphism was first introduced by E. Bessel-Hagen in his work \cite{Bessel-Hagen} and is repeatedly used in the literature (e.g., \cite{Jackiw:1978ar}). It was dubbed ``gauge covariant Lie derivative," (see e.g., \cite{Bourguignon:1992sp, Jacobson:2015uqa}) or ``improved diffeomorphisms'' \cite{Banados:1996yj, Banados:2016zim}. See also  discussions and references in \cite{Prabhu:2015vua}. We use ``augmented diffeomorphism" in order to prevent confusion with a different procedure that is called ``improved energy-momentum tensor" (see \cite{Banados:2016zim,Blaschke:2016ohs,Gieres:2022cpn} to find more on their difference).
	
	In what follows, we use $\eta$ as the augmented diffeomorphism. We  note that the CPSF and the charge variation work  for a generic field-dependent diffeomorphisms or gauge transformations when $\delta\eta \neq 0$ 
	\cite{The-BH-Book}. The above readily generalizes to the cases with several $U(1)$ gauge fields $A^{(a)}$, as $\eta= (\xi, -\xi\cdot A^{(a)})$. {We note that the generalized symmetry generators $\eta=\{\xi, -\xi\cdot A\}$ do not form an algebra, as $[\eta_1, \eta_2]=\{[\xi_1,\xi_2]_{_{\text{L.B}}}, -[\xi_1,\xi_2]_{_{\text{L.B}}}\cdot A+2\xi_1\cdot(\xi_2\cdot F)\}$. Nonetheless, if $\xi_1, \xi_2$ are Killing vectors, $\xi_1\cdot(\xi_2\cdot F)=0$ \cite{Prabhu:2015vua}, and we retain an algebra structure.}

	\section{Gauge invariant black hole charges}
	
	As the first step toward gauge invariant first law, we define gauge invariant black hole charges. Consider a generic stationary axisymmetric black hole solution to \eqref{action}, with stationarity Killing vector (KV) $\xi_\M$ and axisymmetry KVs $\xi_\Ji$. Assume that this black hole has a nondegenerate Killing horizon $\text{H}$ generated by KV $\xi_\H$ and let $\kappa$ and $\Omega^i_\H$, respectively, denote surface gravity and horizon angular velocities. For a stationary black hole, $\xi_\H=\xi_\M-\Omega^i_\H \xi_\Ji$. We can then define the augmented isometries, 
	\begin{align}\label{generators}
		\eta_\S=\frac{2\pi}{\kappa}\{\xi_\H,-\xi_\H\cdot A\},\qquad   \eta_\M=\{\xi_\M,-\xi_\M\cdot A\}, \qquad   \eta_\Ji=\{\xi_\Ji,-\xi_\Ji\cdot A\},
	\end{align}
	where $A$ is the gauge field configuration corresponding to the black hole solution. The above symmetry generators have hence both field dependence and parametric dependence (through $\kappa, \Omega_i$) \cite{HS:2015xlp, The-BH-Book}. In particular, we note that
	\begin{equation}\label{generators-relation}
		\eta_\S=\frac{2\pi}{\kappa} (\eta_\M- \sum_i \ \Omega^i_\H\ \eta_\Ji),
	\end{equation}
	which was already mentioned and used in Sec. 5.4.3 of \cite{The-BH-Book}. 
	
	As usual, we can define the  charge variations as
	\begin{equation}\label{charges-mass}
		\delta M := \oint_{\text{S}_\infty}\boldsymbol{k}_{\eta_\M} (\delta\Phi,\Phi)\,, \qquad \delta J_i := \oint_{\text{S}_\infty}\boldsymbol{k}_{\eta_\Ji} (\delta\Phi,\Phi)\,,\qquad {\delta S} := \oint_{\text{B}}\boldsymbol{k}_{\eta_\S} (\delta\Phi,\Phi)\,,
	\end{equation}
	where  $\text{B}$ is the bifurcation horizon  and $\text{S}_\infty$ denotes a constant time slice at the spatial infinity $i^0$. We note that $M$, $J_i$ are the usual mass and angular momenta and  $S$ is the Wald entropy. These charges as shown in \cite{Wald:1993nt, Iyer:1994ys} are integrable. The virtue of our new definition, using $\eta$ instead of $\xi$, is that these charges are now gauge invariant by construction.
	
	\section{Gauge invariant first law}
	
	Consider the symplectic current $\boldsymbol{\omega}(\delta\Phi, \delta_{\hat{\eta}}\Phi; \Phi)$ where $\hat{\eta}=\{\xi_\H,0\}$.  Integrate the charge expression over a (partial) Cauchy surface $\Sigma$ with two end points (codimension 2 surfaces)  $\text{B}$ and $\text{S}_\infty$, and  recall that $\mrd \boldsymbol{\omega}\approx 0$, to find
	\begin{equation} \label{first law proof 1}
		\int_\Sigma \boldsymbol{\omega}(\delta\Phi, \delta_{\hat{\eta}}\Phi; \Phi) 
		=\oint_{\text{S}_\infty} \boldsymbol{k}_{\hat{\eta}}(\delta\Phi; \Phi) -\oint_\text{B} \boldsymbol{k}_{\hat{\eta}}(\delta\Phi; \Phi).   
	\end{equation}
	Next, recall that
	\begin{subequations}\begin{align}
			\hat{\eta}& =\{\xi_\M,-\xi_\M\cdot A\} - \Omega^i_\H \{\xi_\Ji,-\xi_\Ji \cdot A\}-
			\{0, -\xi_\H\cdot A\}   \\
			&= \eta_\M-\Omega^i_\H \eta_\Ji -
			\{0, -\xi_\H\cdot A\} \label{eta-hat-identities-i0} \\
			&= \frac{\kappa}{2\pi} \eta_\S - \{0, -\xi_\H\cdot A\}, \label{eta-hat-identities-B}
		\end{align}
	\end{subequations}
	where in the last line we used \eqref{generators-relation}. 
	
	{
		We now calculate the left- and right-hand sides of \eqref{first law proof 1}. For the lhs, we note that $\xi_\H$ is a KV; hence, $\delta_{ \hat{\eta}}$ acting on $g_{\mu\nu}$ and the scalar fields $\phi_a$ vanish. Therefore, only the Maxwell part of the action contributes, cf.  \eqref{dH--eq},
		\begin{equation}
			\boldsymbol{\omega}(\delta\Phi, \delta_{\hat{\eta}}\Phi;\Phi)\approx \delta\big((\sigma \star F) \wedge (\mathcal{L}_{\xi_\H} A)\big)- \mathcal{L}_{\xi_\H}\big((\sigma \star F) \wedge \delta A \big).   
		\end{equation}
		Using  $\mathcal{L}_{\xi_\H}(\sigma \star F)=0$ and $\mathcal{L}_{\xi_\H}\delta A=\delta \mathcal{L}_{\xi_\H}A$,  we find  
		\begin{equation}\label{LHS-omega} 
			\int_\Sigma\boldsymbol{\omega}(\delta\Phi, \delta_{\hat{\eta}}\Phi;\Phi) \approx  \int_\Sigma \delta (\sigma \star F) \wedge (\mathcal{L}_{\xi_\H}A).      
		\end{equation}
		The rhs of \eqref{first law proof 1} can be computed using \eqref{eta-hat-identities-i0} and \eqref{eta-hat-identities-B}   
		in the first and second terms respectively,
		\begin{align}
			\oint_{\text{S}_\infty} \boldsymbol{k}_{\hat{\eta}}(\delta\Phi; \Phi) -\oint_\text{B} \boldsymbol{k}_{\hat{\eta}}(\delta\Phi; \Phi) &=    \delta M - \Omega^i_\H \delta J_i -\frac{\kappa}{2\pi}\delta S \nonumber\\ 
			&-\oint_{\text{S}_\infty} \delta \big((\sigma \star F) \wedge (\xi_\H\cdot A)\big)+ \oint_{\text{B}} \delta \big((\sigma \star F) \wedge (\xi_\H\cdot A)\big),\label{RHS-k}
		\end{align}
		where we used {the fact that $\kappa, \Omega^i_\H$ are constants and pulled them out of the integrals} and 
		\begin{align}
			&\boldsymbol{k}_{\{0,-\xi_\H\cdot A\}}(\delta\Phi;\Phi)=\delta \big((\sigma \star F) \wedge (-\xi_\H\cdot A)\big). 
		\end{align}
		Equating \eqref{LHS-omega} and \eqref{RHS-k}, we arrive at the first law
		\begin{equation}\label{first law}
			\delta M= T_\H\delta S +\Omega^i_\H\delta J_i + \slashed{\delta} \Psi,
		\end{equation}
		where $T_\H=\frac{\kappa}{2\pi}$ is the Hawking temperature, and $\slashed{\delta}  \Psi$ is explicitly given as
		\begin{equation}\label{del psi 1}
			\begin{split}
				\slashed{\delta} \Psi &=  \oint_{\text{S}_\infty} \delta (\sigma \star F) \, (\xi_\H \cdot A)-\oint_\text{B} \delta (\sigma \star F) \, (\xi_\H \cdot A)  + \int_\Sigma \delta (\sigma \star F) \wedge (\mathcal{L}_{\xi_\H}A)\\
				&  \approx  \int_\Sigma \delta (\sigma \star F) \wedge ({\xi_\H}\cdot F),
			\end{split}
		\end{equation}
		where in the last equation we used \eqref{EOM} and the identity $\mathcal{L}_{\xi_\H}A=\xi_\H\cdot F + \mrd (\xi_\H\cdot A)$.} (As a comment, recall that ${\xi_\H}\cdot F=\delta_{\eta_\H} A$.) 
	While $\slashed{\delta} \Psi$ is  manifestly gauge invariant, it is not necessarily integrable. Note also that, as we discussed $\delta M, \delta J_i, \delta S$ by construction and $T_\H, \Omega^i_\H$, which are determined only from the metric $g_{\mu\nu}$, are all gauge invariant. So, \eqref{first law} is the promised gauge invariant first law. 
	
	Let us discuss  the $\slashed{\delta} \Psi$ term a bit more. Recall that the electric charge is given by the Gauss law $\delta Q = \oint_{\text{S}_\infty} \delta (\sigma \star F)$, which is obviously gauge invariant. We may then define the gauge invariant electric chemical potential $\mu_{_\text{Q}}$ by 
	\begin{equation}\label{del psi del Q}
		\slashed{\delta} \Psi := \mu_{_\text{Q}} \delta Q. 
	\end{equation}
	So, the first law can be written as
	\begin{equation}\label{first law 2}
		\delta M= T_\H\delta S +\Omega^i_\H\delta J_i + \mu_{_\text{Q}} \delta Q.
	\end{equation}
	While $\mu_{_\text{Q}}$ is gauge invariant, one can present its explicit expression in specific gauges. We will discuss this further in the next section.

	We close this part pointing out that  our analysis and discussions, and hence the result, extend to the cases with any number of $U(1)$ gauge fields [see comments below \eqref{lambda xi.A}] as well as  generic $p$-form gauge fields in any dimension $D$ \cite{Compere:2007} and, in particular, to the case of a cosmological constant if it is viewed as the charge associated with a $D-1$-form field \cite{Chernyavsky,Hajian:2021hje}. 
	
	\section{Zeroth law}\label{sec:zeroth}
	
	To complete the gauge invariant description of black hole thermodynamics, we also need to discuss the zeroth law. The zeroth law is a statement about a system to be  in thermal equilibrium, when the flow of charges is zero. Since the flow of charges is proportional to the gradient of the corresponding chemical potential, the zeroth law implies the  constancy of the chemical potentials in the thermal system. With the same token, the zeroth law of black hole thermodynamics requires   constancy of the temperature and other chemical potentials on the horizon. In our case, that is constancy of $T_\H, \Omega^i_\H, \mu_{_\text{Q}}$. There are established textbook arguments proving constancy of these for stationary black holes with a Killing horizon; see chapter 5 of \cite{The-BH-Book} {and \cite{Ghosh:2020dkk} for analysis in beyond Einstein gravity theories.} Here, we only focus on the interplay of gauge invariance and the constancy of chemical potentials. Since  $T_\H, \Omega^i_\H$ and the usual proofs for their constancy  are manifestly gauge invariant, we only discuss the case of $\mu_{_\text{Q}}$. 
	
	Since we have already established the gauge invariance of $\mu_{_\text{Q}}$, to argue for its constancy over the horizon, without loss of generality, we can work in a specific gauge. To this end, note that \eqref{del psi 1} is not localized on the horizon; in particular, it has a term which is an integral over $\Sigma$. Note also that 
	since $\xi_\H$ is a KV and a symmetry of the black hole solution, $\mathcal{L}_{\xi_\H}F=0$.  We now show that there exists a gauge in which $\mathcal{L}_{\xi_\H}A=0$. First, we note that $\mathcal{L}_{\xi_\H}F=0$ implies  $\mathcal{L}_{\xi_\H}\mrd A=\mrd \mathcal{L}_{\xi_\H} A=0$. Then, by the Poincar\'e Lemma $\mathcal{L}_{\xi_\H} A=\mrd \lambda$ for a scalar $\lambda$ and if we take $\lambda=\mathcal{L}_{\xi_\H} \tilde\lambda$, which we can always do so, $\mathcal{L}_{\xi_\H} \tilde A=0$  in the $\tilde A= A-\mrd \tilde\lambda$ gauge.  In this gauge, \eqref{del psi 1} reduces to integrals over the boundaries,
	\begin{equation}\label{Psi-mu-1}
		\slashed{\delta} \Psi =  \oint_{\text{S}_\infty} \Phi{_{\text{S}_\infty}}\ \delta (\sigma \star F)  +\oint_\text{B} \Phi_\H\ \delta (\sigma \star F),\qquad  \Phi_\H= \xi_\H\cdot A|_{_{\text{B}}}, \quad \Phi_{_{\text{S}_\infty}}=\xi_\H\cdot A|_{_{\text{S}_\infty}}.
	\end{equation}
	
	We now show the constancy of $\Phi_\H$  over the horizon and $\Phi_{_{\text{S}_\infty}}$ over ${{\text{S}_\infty}}$. In the $\mathcal{L}_{\xi_\H} A=0$ gauge, constancy of $\Phi_\H$ over the horizon amounts to showing $\zeta^\alpha\partial_\alpha \Phi_\H=0$ for any horizon tangent vector $\zeta$. To see the latter, recall that  horizon is a null hypsersurface  generated by the orbits of $\xi_\H$, and for $\zeta=\xi_\H$, we have $\xi_\H^\alpha\partial_\alpha \Phi_\H=\mathcal{L}_{\xi_\H} \Phi_\H=0$. It hence remains to show $\zeta^\alpha\partial_\alpha \Phi_\H=0$ for any vector $\zeta$ tangent to the bifurcation surface B, $(\zeta\cdot \xi_\H)_{_\text{B}}=0$:
	\begin{equation}
		\zeta^\alpha\partial_\alpha \Phi_\H 
		=\zeta\cdot \mrd (\xi_\H\cdot A) \mid_{_\text{B}}
		=\zeta\cdot(\mathcal{L}_{\xi_\H}A-\xi_\H\cdot \mrd A)\mid_{_\text{B}}=-\zeta\cdot(\xi_\H\cdot F)\mid_{_\text{B}}=0,   
	\end{equation}
	where in the second equality, we used the Cartan identity; the third equality follows from the gauge choice $\mathcal{L}_{\xi_\H} A=0$, and the last equality is a result of the vanishing of $\xi_\H$ on the bifurcation horizon B where the field strength $F$ has finite components.  {In fact, using  gauge field equation \eqref{EOM}, one can show that $\xi_\H\cdot F\mid_{_\text{B}}\approx \mrd X$ for a gauge invariant scalar function $X$ defined at B. There can't be any such function $X$, and hence $\xi_\H\cdot F\mid_{_\text{B}}\approx 0$; see \cite{Prabhu:2015vua} for more formal proof. (As a side comment, recalling \eqref{delta-eta-A}, that is $\delta_{\eta_\H} A\mid_{_\text{B}}\approx0$.)} Therefore, $\Phi_\H$ is  constant over the horizon H. A similar analysis also shows that $\Phi_{_{\text{S}_\infty}}$ is a constant over ${\text{S}_\infty}$. Therefore, we may rewrite  \eqref{Psi-mu-1} as
	\begin{equation}\label{Psi-mu}
		\slashed{\delta} \Psi=(\Phi_\H-\Phi{_{\text{S}_\infty}})\delta Q,  
	\end{equation}
	where we used $Q= \oint_{\text{S}_\infty} (\sigma \star F)=-\oint_{\text B} (\sigma \star F)$. With the above, we recover the standard term in the first law.

	\section{Concluding remarks}
	
	We have given an explicitly gauge invariant derivation of  the zeroth and first laws of black hole thermodynamics. We achieved this by extending the notion of Killing vectors to symmetries that include certain field-dependent gauge transformations. At the heart of our proposal is the gauge invariant transformation for the gauge field \eqref{delta-eta-A}; {our analysis here completes the discussions in \cite{Prabhu:2015vua,HS:2015xlp}}. Despite the fact that we limited our presentation to a specific class of theories, as our arguments are within the covariant phase space formalism (CPSF), our analyses and results hold for any diffeomorphism and gauge invariant theory. In particular our analysis extends to cases with higher $p$ forms and/or asymptotic (A)dS black holes. 
	
	It is well known that in the context of CPSF there are certain ``ambiguities'' in the definition of charge variations \cite{Iyer:1994ys}. Also, there are  ambiguities in Wald's definition of the entropy \cite{Wald:1993nt,Wald:1999wa}. {These ambiguities are ultimately defining the boundary theory, its symplectic form and its Lagrangian \cite{Adami:2022ktn}; here, the boundary theory resides at H or $i^0$. Therefore, the gauge invariance of the full theory, which is our starting point, implies gauge invariance of these ambiguities. }
	
	CPSF leads to charge variations. While the first law deals with charge variations, we need to make sure the charges are well defined; i.e., they are integrable over the solution space. It has been argued that the charge integrability in general depends on the slicing of the solution phase space and can be achieved with appropriate symmetry generators \cite{Adami:2020ugu,Adami:2021nnf}. For the case of our interest here, stationary black holes and their thermodynamical laws, we need to make sure of the integrability of charges under parametric variations, variations in parameter space of black hole solutions \cite{HS:2015xlp, The-BH-Book}. One can show that our augmented symmetry generators $\eta$'s yield integrable charges, while the $\xi$ symmetry generators do not; see the Appendix for an explicit example.
	
	Besides the zeroth and first laws, one may naturally ask about gauge (non)invariance of the usual derivations of second and third laws of black hole thermodynamics. Hawking's area theorem \cite{Bardeen:1973gs} (and its extensions to generic modified gravity theories) provides the basis for  generalized second law arguments; see e.g., \cite{Bekenstein:1974ax, Wall:2009wm}. If the notion of the entropy (of black hole and the surrounding systems) are gauge invariant, they lead to the gauge invariant notion of the generalized second law. Here, we have provided such a gauge invariant notion for the entropy. The most rigorous derivation of the third law, to our knowledge, is due to Israel \cite{Israel:1986gqz}. If the thermodynamical quantities entering the analysis are gauge invariant, which here we established they are, the arguments in \cite{Israel:1986gqz} seem to be gauge invariant. Establishing these more explicitly and in some clarifying examples is a well worth exercise to take.

	Here, we limited our analysis to stationary black holes with a bifurcate Killing horizon, while the laws of black hole thermodynamics can hold in general gravitational systems where some of these conditions are relaxed or replaced by less restrictive assumptions. For instance, it is known that the near-horizon-extremal geometries that are not black holes in the usual sense also obey laws of thermodynamics, albeit at zero temperature \cite{Hajian:2013lna}. One may check that our analysis here readily extends to such cases as well as to extremal black holes, which do not have a bifurcate Killing horizon. It is also interesting to study gauge invariance by relaxing stationarity or explore in more detail black holes in asymptotically (A)dS cases or in more general theories like those discussed in \cite{Hajian:2020dcq}.

	\noindent \textbf{Acknowledgements:} We would like to thank Kartik Prabhu for detailed discussions on his related work \cite{Prabhu:2015vua} and Ibrahim Shehzad for comments. This work has been supported by TÜBITAK international researchers program No. 2221. M.M.Sh.J acknowledges partial support by  SarAmadan Grant No ISEF/M/401332. 
	
	\appendix
	

	\section{An example: Kerr-Newman black hole}\label{sec:example}
	
	To clarify more the abstract analysis in the main text, we present an example for which we  also discuss the integrability of charges which we skipped in the general discussions. As our example we choose the Kerr-Newman black hole solution to the four-dimensional Einstein-Maxwell theory 
	\begin{equation}
		\mathcal{L}=\frac{1}{16\pi}(R-F^2),    
	\end{equation}
	and calculate $\slashed{\delta} Q_\eta$ in \eqref{del H}. For an arbitrary $\eta=\{\xi,\lambda\}$, $\boldsymbol{k}_{\eta}$ is found (see \cite{Ghodrati:2016vvf} for the details) \begin{equation}\label{k EH integrand}
		\begin{split}
			k_{\eta}{}^{\mu\nu}(\delta g, g) &=\dfrac{1}{16 \pi}\Big(\Big[\xi^\nu\nabla^\mu h-\xi^\nu\nabla_\alpha h^{\mu\alpha}+\xi_\alpha\nabla^{\nu}h^{\mu\alpha}+\frac{1}{2}h\nabla^{\nu} \xi^{\mu}-h^{\alpha\nu}\nabla_\alpha\xi^{\mu}\Big]-[\mu\leftrightarrow\nu]\Big)\\
			&+\frac{1}{4\pi}\Big(\big(-\frac{h}{2} F^{\mu\nu}\!-F^{\rho[\mu}h^{\nu]}_{\;\;\rho}-\delta F^{\mu\nu}\big)({\xi}^\alpha A_\alpha+\lambda)-\ F^{\mu\nu}\xi^\alpha \delta A_\alpha-\,F^{\alpha[\mu}\xi^{\nu]} \delta A_\alpha\Big),
		\end{split}    
	\end{equation}
	in which $h^{\mu\nu}:= g^{\mu \beta}g^{\nu\rho}\delta g_{\beta \rho}$ and $h:= h^\mu_{\,\,\mu}$. Note that ${k}_{\eta}{}^{\alpha\beta}:=\epsilon^{\mu\nu\alpha\beta}\ (\boldsymbol{k}_{\eta})_{\mu\nu}$, and  the charge variation is diffeomorphism invariant only if it acts also on the $\eta$ (see more details in e.g. \cite{Altas:2018pkl,Altas:2018zjr}). 
	
	The Kerr-Newman (KN) metric and the gauge field in coordinates $x^\mu=(t,r,\theta,\varphi)$ and in an arbitrary gauge is \cite{Newman:1965my}
	\begin{equation}\label{KN gauge}
		\begin{split}
			\mathrm{d}s^2 &= -(1-f)\mathrm{d}t^2+\frac{\rho ^2}{\Delta_r}\mathrm{d}r^2+{\rho ^2} \mathrm{d}\theta ^2 -2 fa\sin ^2 \theta\,\mathrm{d}t \mathrm{d}\varphi+\left( {r^2+a^2}+fa^2\sin ^2\theta \right)\sin ^2\theta\,\mathrm{d}\varphi ^2\,, \\
			A &=\frac{qr}{\rho^2}(\mrd t-a\sin^2 \theta \,\mrd \varphi)+ \mrd \tilde\lambda, 
		\end{split}
	\end{equation}
	where $m$, $a$, and $q$ are free parameters, \begin{align}
		\rho^2 &:= r^2+a^2 \cos^2 \theta\,,\qquad \Delta_r := r^2+a^2-2mr + q^2\,,\qquad
		f:=\frac{2mr-q^2}{\rho ^2}\,,
	\end{align}
	and $\tilde\lambda=\tilde\lambda (t,r,\theta,\varphi)$ is an arbitrary scalar function. Moreover
	\begin{equation}
		\begin{split}
			r_\H=m+\sqrt{m^2-a^2-q^2}, \quad \kappa &= \frac{{r_\H^2}-{a^2}}{2r{_\H}(r_\H^2+a^2)},   \quad \Omega_\H=\frac{a}{r_\H^2+a^2},\qquad \Phi_\H= \frac{q r_\H}{r^2_\H+a^2}\\
			\xi_\M=\partial_t,\qquad \xi_{_J} &=-\partial_\varphi, \qquad \xi_{_S}=\frac{2\pi}{\kappa}(\xi_\M-\Omega_\H\xi_{_J}). 
		\end{split}
	\end{equation}
	
	The first issue we want to check by this example is whether $\eta_\M$, $\eta_{_J}$, and $\eta_\S$ in \eqref{generators} generate the mass, angular momentum, and entropy respectively. To this end, as discussed in \cite{HS:2015xlp}, we need to consider the parametric variations. That is, consider set of KN solutions \eqref{KN gauge} and vary the solution parameters $m,a,q$: 
	\begin{equation}\label{parametric var}
		\hat{\delta}g_{\mu\nu}=\frac{\partial g_{\mu\nu}}{\partial m}\delta m +\frac{\partial g_{\mu\nu}}{\partial a}\delta a  + \frac{\partial g_{\mu\nu}}{\partial q}\delta q, \qquad \hat{\delta} A_\mu= \frac{\partial A_{\mu}}{\partial m}\delta m + \frac{\partial A_{\mu}}{\partial a}\delta a+ \frac{\partial A_{\mu}}{\partial q}\delta q.   
	\end{equation}
	After performing the calculation using these variations in  
	\begin{equation}
		\hat{\slashed{\delta}} Q_{\eta_\M}=\oint_{\text{S}_\infty} \mrd\theta\mrd \varphi\  \sqrt{-g}\  k_{\eta_\M}^{tr}, \quad \hat{\slashed{\delta}} Q_{\eta_{_J}}=\oint_{\text{S}_\infty} \mrd\theta\mrd \varphi\  \sqrt{-g}\  k_{\eta_{_J}}^{tr}, \quad  \hat{\slashed{\delta}} Q_{\eta_\S}=\oint_\text{B} \mrd\theta\mrd \varphi\  \sqrt{-g}\  k_{\eta_\S}^{tr},   
	\end{equation}
	we find 
	\begin{equation}
		\hat{\delta} Q_{\eta_\M}=\delta m, \qquad \hat{\delta} Q_{\eta_{_J}}=\delta m a+ a\delta m, \qquad \hat{\delta} Q_{\eta_{_S}}=\frac{2\pi}{\kappa}\big(\delta m -\Omega_\H (m\delta a+a\delta m) - (\frac{qr_\H}{r_\H^2+a^2}) \delta q\big),    
	\end{equation}
	which are manifestly integrable \emph{if} we take $m,a,q$ to be independent parameters and produce the standard charges 
	\begin{equation}
		M:=Q_{\eta_\M}=m, \qquad J:=Q_{\eta_{_J}}=ma, \qquad S={\pi(r_{_\mathrm{H}}^2+a^2)},     
	\end{equation}
	respectively. The result is independent of $\tilde\lambda$, confirming that the augmented generators in \eqref{generators} produce gauge invariant charges. 
	
	The second issue to investigate is the gauge noninvariance and nonintegrability of the charge variations associated to $\xi_\M$, $\xi_{_J}$ and $\xi_{_S}$ and show that the charge variations depend on  $\tilde\lambda$ and are not integrable. To this end, it suffices to focus on a subset of $\tilde\lambda$, which produces finite and coordinate-independent charge variations. 
	
	Let us begin with $\xi_\M$ and  consider $\tilde\lambda=\tilde\lambda(t,r,\theta)
	$ such that at $\text{S}_\infty$, it tends to a  linear function in time $\tilde\lambda_\infty(m,a,q) \ t$ \cite{Gao:2003ys}. If we insert $\{\partial_t,0 \}$ and the parametric variations \eqref{parametric var} into  \eqref{k EH integrand},  we find 
	\begin{equation}\label{M nonintegrable}
		\slashed{\delta} Q_{\xi_\M}=\delta m +\tilde\lambda_\infty(m,a,q) \delta q.  
	\end{equation}
	Presence of $\tilde\lambda$ in the result clearly shows  its gauge dependence. Moreover, it is not integrable for arbitrary $\tilde\lambda_\infty$; i.e., it is not $\delta M(m,a,q)$ for a function $M(m,a,q)$.
	
	To check the gauge noninvariance of the charge variation of $\xi_{_J}$, we can choose $\tilde\lambda=\tilde\lambda(r,\theta,\varphi)$ such that at $\text{S}_\infty$, it tends to a constant $\tilde\lambda_\infty(m,a,q)$ times $\varphi$. A similar calculation shows 
	\begin{equation}\label{J nonintegrable}
		\slashed{\delta} Q_{\xi_{_J}}=\delta(ma - q\tilde\lambda_\infty),
	\end{equation}
	while integrable, it  is gauge dependent for arbitrary $\tilde\lambda_\infty$. 
	
	To show the nonintegrability of $\xi_\S=\frac{2\pi}{\kappa}\xi_\H$, one can follow similar steps while integrating over B instead of $\text{S}_\infty$. However, there is a shortcut which we follow here. The point is that even for the case of $\tilde\lambda=0$,  $\xi_\S$ has nonintegrable charge variation. This important result was first reported in \cite{HS:2015xlp}, and we summarize it here. Let us set $\tilde{\lambda}=0$ in \eqref{KN gauge}. In this gauge, the KVs $\xi_\M$ and $\xi_{_J}$ are exact symmetries, and $\boldsymbol{\omega}$ in \eqref{first law proof 1}  vanishes for these. So by Stokes' theorem, their charge variations can be calculated not only at $\text{S}_\infty$ but also on any other smooth closed codimension-2 surface that encompasses  B (including the B itself) producing the same result.  Therefore, by $\xi_{_S}=\frac{2\pi}{\kappa}(\xi_\M-\Omega_\H\xi_{_J})$ and by setting $\tilde{\lambda}=0$ in \eqref{M nonintegrable} and \eqref{J nonintegrable}, we find  
	\begin{equation}\label{S nonintegrable}
		\slashed{\delta} Q_{\xi_{_S}}=\frac{2\pi}{\kappa}\big(\delta m -\Omega_\H (m\delta a+a\delta m)\big),  
	\end{equation}
	which is not integrable over the whole set of parameters $m,a,q$.

\end{document}